\documentclass[runningheads]{llncs}
\usepackage[T1]{fontenc}
\usepackage{atbegshi}
\usepackage{tikzpagenodes}
\usepackage{tikz}
\usepackage{graphicx}
\usepackage[
    natbib=true,
    style=numeric,
    sorting=none
]{biblatex}
\begin{document}
\AtBeginShipout{%
  \AtBeginShipoutUpperLeft{%
    \begin{tikzpicture}[remember picture,overlay]
      \node[
        anchor=north,
        yshift=-6mm,
        text width=\paperwidth-2cm,
        align=center,
        font=\footnotesize
      ] at (current page.north) {%
        This version of the article has been accepted for publication in Springer CCIS, volume 2775, after peer review (when applicable) but is not the Version of Record and does not reflect post-acceptance improvements, or any corrections. The Version of Record is available online at springnature. Use of this Accepted Version is subject to the publisher’s Accepted Manuscript terms of use https://www.springernature.com/gp/open-research/policies/accepted-manuscriptterms.
      };
    \end{tikzpicture}%
  }%
}

\title{The Evolution of Reranking Models in Information Retrieval: From Heuristic Methods to Large Language Models}
\titlerunning{The Evolution of Reranking Models in IR}
\author{Tejul Pandit\inst{1}\orcidID{0009-0006-4376-1063} \and
Sakshi Mahendru\inst{1}$^{*}$ \and
Meet Raval\inst{2}$^{*}$ \and
Dhvani Upadhyay\inst{3}}
\authorrunning{T. Pandit et al.}
\institute{Palo Alto Networks, USA \and
University of Southern California, USA \and
Dhirubhai Ambani University, India\\
\email{tejulpandit96@gmail.com}}

\maketitle              % typeset the header of the contribution
% \begin{center}
% \footnotesize
% \textit{This version of the article has been accepted for publication in Springer CCIS, Vol.~2775.
% It is not the Version of Record and does not reflect post-acceptance changes or corrections.
% The final Version of Record is available at Springer Nature.
% Use of this Accepted Manuscript is subject to the publisher’s terms.}
% \end{center}
% \vspace{1em}

\begingroup
\renewcommand\thefootnote{*}
\footnotetext{Equal contribution}
\endgroup

\begin{abstract}
Reranking is a critical stage in contemporary information retrieval (IR) systems, improving the relevance of the user-presented final results by honing initial candidate sets. This paper is a thorough guide to examine the changing reranker landscape and offer a clear view of the advancements made in reranking methods. We present a comprehensive survey of reranking models employed in IR, particularly within modern Retrieval Augmented Generation (RAG) pipelines, where retrieved documents notably influence output quality.

We embark on a chronological journey through the historical trajectory of reranking techniques, starting with foundational approaches, before exploring the wide range of sophisticated neural network architectures such as cross-encoders, sequence-generation models like T5, and Graph Neural Networks (GNNs) utilized for structural information. Recognizing the computational cost of advancing neural rerankers, we analyze techniques for enhancing efficiency, notably knowledge distillation for creating competitive, lighter alternatives. Furthermore, we map the emerging territory of integrating Large Language Models (LLMs) in reranking, examining novel prompting strategies and fine-tuning tactics. This survey seeks to elucidate the fundamental ideas, relative effectiveness, computational features, and real-world trade-offs of various reranking strategies. The survey provides a structured synthesis of the diverse reranking paradigms, highlighting their underlying principles and comparative strengths and weaknesses.

\end{abstract}
\section{Introduction}
\begin{figure}
\begin{center}
\includegraphics[width=\textwidth]{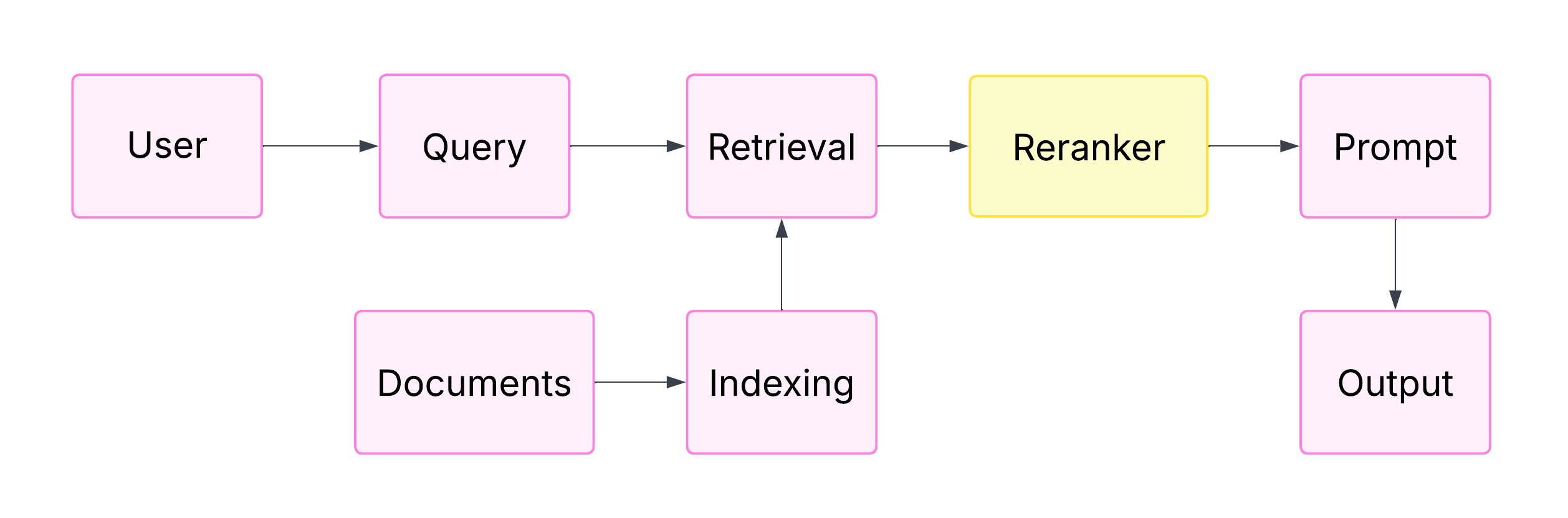}
\end{center}
\caption{RAG approach highlighting a post-retrieval step of Reranking documents.} \label{fig1}
\end{figure}

Information retrieval (IR) systems are essential in today's digital world, with their ability to power anything from online search and recommender systems to sophisticated question-answering and knowledge management platforms. A critical component within these systems is the reranking stage. Reranking carefully reorders these candidates to show the user the most pertinent results after an initial, often rapid, retrieval phase that produces an extensive collection of candidate documents. This dramatically improves the overall quality and efficacy of the IR system\cite{RerankZilliz}. Notably, the significant development in Large Language Models (LLMs) has subsequently driven substantial improvements and interest in Retrieval Augmented Generation (RAG) pipelines, where the precision of retrieved context, refined by reranking, is crucial for generating accurate and relevant outputs\cite{Pinecone}. Based on \cite{gao2024retrievalaugmentedgenerationlargelanguage}, we represent the typical placement of the reranking component within a RAG pipeline in Figure \ref{fig1}.

\vspace{-1.5em}

Reranking methods have evolved significantly, from heuristic scoring and traditional learning-to-rank models to deep learning and, more recently, LLMs, enabling richer semantic understanding and inter-document relationships.

This survey provides a comprehensive exploration of the evolving field of reranking models in IR. We trace the development trajectory from classical learning-to-rank methods in Section \ref{ltr-rerankers}, followed by Deep learning rerankers in Section \ref{dl-rerankers}. We also cover methods aimed at improving the efficiency of reranking models, with a particular focus on knowledge distillation techniques in Section \ref{kd-rerankers}. Finally, we examine the cutting-edge integration of LLM-based rerankers in Section \ref{llm-rerankers}. This study aims to provide scholars and practitioners with a valuable guide for navigating this ever-evolving subject by compiling significant advancements and empirical findings throughout the reranking timeline.

\section{Background}
Ranking approaches, which form the basis for many reranking techniques, can broadly be categorized based on their supervision strategies and how they process documents during training and scoring: pointwise, pairwise, and listwise. 

\textbf{Pointwise approaches} consider a single document at a time within the loss function. They train a classifier or regressor to predict the relevance score of an individual document for a given query. The final ranking is then produced by sorting documents based on these predicted scores.

\textbf{Pairwise approaches} focus on pairs of documents. Given a pair of documents, the model aims to determine the optimal relative ordering between them, comparing this prediction against the ground truth.

\textbf{Listwise approaches} directly consider and optimize the ranking of an entire list of documents. These methods aim to produce the optimal ordering for the complete set of retrieved candidates.

\section{Learning to Rank in Reranking Models} \label{ltr-rerankers}
Learning drastically changed the reranking phase of contemporary information retrieval (IR) systems to Rank (LTR). This section of the survey charts LTR's evolutionary trajectory in reranking, focusing on methodological shifts, core problems, and model sophistication, from foundational probabilistic methods to pairwise learning, listwise optimization, and introduction to deep learning models.

Early LTR (late 1980s- early 1990s) applied statistical methods to estimate relevance. Key efforts included \cite{10.1145/65943.65944} using polynomial regression for direct relevance probability estimation, \cite{10.1145/133160.133199} with multi-stage logistic regression for composite clues, and \cite{10.5555/188490.188554} learning optimal linear combinations of expert scores, presaging meta-search. These established data-driven rankings' viability.

The late 1990s- early 2000s saw pairwise learning ascend, driven by machine learning advances. Friedman's \cite{4a848dd1-54e3-3c3c-83c3-04977ded2e71} provided the Gradient Boosted Decision Tree (GBDT) foundation crucial for later LTR. Algorithms like \cite{10.5555/945365.964285} applied boosting to learn from pairwise preferences, while \cite{10.1145/1102351.1102363} introduced influential neural networks for pairwise ranking by minimizing cross-entropy between predicted and target pair probabilities.

Limitations of generic pairwise losses spurred refinement towards IR-specific metrics. \cite{10.1145/1148170.1148205} Adapted Ranking SVM by weighting misranked pairs based on IR relevance. The pivotal \cite{paper8_ltr} enabled direct optimization of non-smooth IR metrics (like NDCG) by defining gradients ($\lambda$-values) based on metric changes from swaps, guiding pairwise models to optimize listwise objectives. Other works like \cite{paper9_ltr} and \cite{10.1145/1277741.1277790} also explored pairwise regression for preferences and direct MAP optimization.

This led to explicit listwise approaches, treating entire document lists as training instances. \cite{cao2007learning} pioneered a listwise loss based on top-one probabilities. Others like \cite{QIN2008838} and \cite{tsai2006frank} (both 2007) proposed query-level losses. Boosting frameworks like \cite{10.1145/1277741.1277809} were adapted to optimize IR metrics directly. Probabilistic list models, such as \cite{10.1145/1390156.1390306} using the Plackett-Luce model, gained traction. The highly successful \cite{burges2008ranking} integrated MART (GBDTs) with LambdaRank's gradients, becoming a state-of-the-art baseline. The scalability of such tree-based models was later significantly enhanced by systems like \cite{Chen_2016}.

From the mid-2010s, deep learning (DL) began to dominate LTR, offering powerful representation learning and contextual modeling. Using techniques like \cite{Schroff_2015} triplet loss for metric learning, neural IR models began learning potent query-document embeddings. Contextual DL models, using RNNs or Transformers (e.g., \cite{10.1145/3209978.3209985}, \cite{10.1145/3397271.3401104}, \cite{pei2019personalizedrerankingrecommendation}), excelled at capturing document interdependencies within lists. Simultaneously, DL spurred innovations in preference modeling (e.g., \cite{köppel2019pairwiselearningrankneural}) and end-to-end metric optimization using differentiable surrogates or black-box methods (e.g., \cite{8953684}, \cite{swezey2021pirankscalablelearningrank}). We elaborate on more sophisticated DL-based rerankers in Section \ref{dl-rerankers}.

Alongside these mainstream trends, diverse strategies continued to enrich LTR. Evolutionary algorithms \cite{paper25_ltr}, \cite{10.1145/3019612.3019696} directly optimized IR metrics. Other approaches included ordinal classification \cite{li2008learning}, incorporating preference magnitudes \cite{10.1145/1645953.1646057}, and hybrid models combining objectives or data sources \cite{10.1145/1835804.1835928}.

\section{Deep Learning Techniques for Reranking} \label{dl-rerankers}
Deep learning models have significantly advanced the reranking task by capturing intricate semantic relationships between queries and documents.

\subsection{Transformer-based Rerankers (BERT and T5)}
Transformer architectures dominate recent reranking advancements, notably BERT-style encoders and T5-style sequence-to-sequence models.
\subsubsection{BERT-like Cross-Encoders}
Cross-encoders based on BERT\cite{devlin2019bertpretrainingdeepbidirectional} and its variants jointly encode query-document pairs, enabling rich token-level interactions via self-attention\cite{vaswani2023attentionneed}. They output a relevance score.

Training pipelines are crucial. \cite{nogueira2019multistagedocumentrankingbert} uses sequential sparse retrieval (BM25), pointwise BERT reranking, and pairwise BERT reranking. Integrated pipelines like \cite{glass2022re2gretrievererankgenerate} combine retrieval, reranking, and generation using a BERT-based model trained on MS MARCO\cite{bajaj2018msmarcohumangenerated}.

Efficiency is a key challenge. \cite{khattab2020colbertefficienteffectivepassage} offers a scalable alternative using contextualized late interaction. It encodes queries and documents independently with BERT, preserving token embeddings. Similarity is calculated via Maximum Similarity (MaxSim), summing the max dot products between query and document tokens. This allows document precomputation, drastically reducing latency while maintaining accuracy. Adaptation for multilingual scenarios uses methods like Adapters or SFTMs\cite{litschko2022parameterefficientneuralrerankingcrosslingual} with mBERT, decoupling language and task adaptation to mitigate the "curse of multilinguality"\cite{chang2023multilingualitycurselanguagemodeling}.

\subsubsection{T5-based Sequence-to-Sequence Rerankers}
Seq2seq models like T5\cite{raffel2023exploringlimitstransferlearning} offer an alternative generative approach. They frame relevance prediction as a text-to-text task, leveraging pretraining benefits.

The seminal work \cite{nogueira2020documentrankingpretrainedsequencetosequence} fine-tuned T5 to generate "true" or "false" given a query-document pair, using the probability of generating "true" as the relevance score. \cite{zhuang2022rankt5finetuningt5text} proposes directly outputting numerical scores using the T5 encoder-decoder or encoder-only architecture, supporting fine-tuning with standard ranking losses. Listwise losses often yield better performance. \cite{ju2022texttotextmultiviewlearningpassage} can improve generalization by jointly training on ranking and auxiliary tasks like passage-to-query generation. \cite{yoon2024listt5listwisererankingfusionindecoder} adapts T5 for listwise reranking using the Fusion-in-Decoder (FiD) architecture and an efficient m-ary tournament sort inference, mitigating positional biases seen in other LLM-based listwise methods\cite{liu2023lostmiddlelanguagemodels}.

\subsection{Advanced Interaction and Alternative Architectures}

Research also explores models that explicitly model interactions between multiple candidates or use different architectural foundations.

Methods modeling joint comparison between candidates, like \cite{song2024comparingneighborsmakeseasy}, use a Transformer encoder to jointly encode the query and multiple candidates, capturing inter-candidate relationships efficiently within multi-stage pipelines (e.g., BE-CMC or BE-CMC-CE). \cite{liu2025listconrankercontrastivetextreranker} uses a specialized ListAttention mechanism allowing passages to attend to each other, combined with Circle Loss\cite{sun2020circlelossunifiedperspective} for improved contrastive learning.

Beyond Transformers, other models are explored. \cite{dong2024dontforgetconnectimproving} builds a document graph from retrieved documents, using Graph Convolutional Networks\cite{paper25_dl} to update representations based on relationships derived from Abstract Meaning Representation (AMR). \cite{xu2024rankmambabenchmarkingmambasdocument} evaluates the Mamba architecture\cite{gu2024mambalineartimesequencemodeling} for document reranking, showing competitive performance against Transformers but highlighting current implementation speed challenges.

\section{Efficiency Techniques: Knowledge Distillation} \label{kd-rerankers}
Modern search engines and RAG systems have demonstrated considerable efficacy in question answering. However, their performance often encounters limitations when confronted with complex, knowledge-intensive tasks that demand nuanced reasoning. Such tasks, including the identification of coding/configuration errors, medical diagnosis, financial forecasting, and strategic marketing approach formulation based on case studies, necessitate more than superficial relevance matching, domain-specific knowledge.

\subsection{Distillation Training Strategies for Reasoning-Aware Reranking}
Based on our research across recent papers, the methodologies for distilling reasoning capabilities into smaller LMs for reranking can be broadly classified based on the nature of the supervisory signal transferred from the teacher to the student model. 

We delineate two primary categories - Label Distillation \& Rationale-Enhanced Distillation.
% \begin{itemize}
%     \item Label Distillation
%     \item Rationale-Enhanced Distillation
% \end{itemize}

\cite{kang2023knowledgeaugmentedreasoningdistillationsmall} claims a KARD-trained 250M T5 model surpassed a fine-tuned 3B T5 model and proposes a unique distillation process to transfer the learning from LLMs to small LM using supervised fine tuning process. It follows a two-stage training process (training the small language model (LM) with knowledge, and training a helper "reranker") followed by an inference process that uses both these trained components to answer new questions effectively.

In the first stage, we train the reranker via a teacher LLM that generates rationales for ground truth questions and documents, making the reranker "rationale-aware" or "reasoning-aware," bridging the gap between optimal retrieval for rationale generation and practical inference-time retrieval. In the next stage, small LLM is trained using explicit knowledge augmentation in addition to query and rational, which is theoretically motivated to reduce the memorization burden on small LMs. \cite{jia2024bridgingrelevancereasoningrationale} presents novel frameworks like RADIO (RAtionale DIstillation) to mitigate the  "preference misalignment" gap in the RAG pipeline. This gap can result in the generator receiving documents that, while generally relevant, lack the precise information needed for high-quality, reasoned responses. Recent research has begun to address this by focusing on how the reasoning process itself can inform document selection. It involves fine-tuning the reranker component, denoted as R\_$\theta$ and typically instantiated as a powerful cross-encoder architecture, utilizing a contrastive learning objective.

\cite{ji2025reasoningrankteachingstudentmodels} and \cite{samarinas2025distillationrefinementreasoningsmall} addresses the lack of explainability and transparency in neural document reranking models. It introduces a novel framework that separates and distills both direct relevance and comparative reasoning from a LLM teacher into LLaMA. Both the models were evaluated across a wider range of datasets.

The seminal work \cite{hinton2015distillingknowledgeneuralnetwork} introduced the concept of knowledge distillation, using "soft targets" (probabilities from the teacher's softmax, often with a temperature parameter T) and a modified cross-entropy loss (or matching logits, which it shows is a special case) to transfer knowledge. It defines the basic distillation process and loss. However, this standard approach can suffer from over-calibration, where the student too rigidly imitates an imperfect teacher. \cite{10.1145/3578337.3605120} proposes an improved distillation process by integrating an alternative contrastive loss (LBKL) with the standard Kullback-Leibler (KL) Divergence loss\cite{10.1214/aop/1176996454}. This LBKL allows the student model to more conservatively and adaptively learn from the teacher, permitting deviation when the teacher's guidance might be suboptimal for certain examples. The strategy aims to balance the benefits of teacher imitation with an explicit contrastive objective, demonstrably improving re-ranking performance by making the knowledge transfer more robust to teacher errors.

\subsection{Comparison \& Observations}

\cite{kang2023knowledgeaugmentedreasoningdistillationsmall}'s explicit knowledge augmentation strategy might be advantageous for tasks requiring complex knowledge synthesis beyond document content. \cite{ji2025reasoningrankteachingstudentmodels} is specifically optimized for generating structured, interpretable rationales and performing precise reranking of a given set of documents within a single, efficient student model. For pure document reranking with explainability, the authors found unified distillation particularly effective. \cite{samarinas2025distillationrefinementreasoningsmall} extends \cite{ji2025reasoningrankteachingstudentmodels} distilling approach on web-scraped Q\&A to specifically refine the student's ability to generate these step-by-step explanations before relevance judgment.

The development of such models is crucial for democratizing access to sophisticated reasoning capabilities, overcoming the deployment barriers of LLMs, and paving the way for more transparent and trustworthy IR systems. While not a distillation method itself, frameworks like Rank1\cite{weller2025rank1testtimecomputereranking}, which focus on optimizing test-time computation for reranking, stand to benefit significantly from highly efficient yet effective distilled models that can perform sophisticated reasoning within tight latency budgets.

\section{LLM Based Rerankers} \label{llm-rerankers}

LLM-based rerankers are effective in listwise reranking settings where the goal is to produce an ordered document list without explicit relevance scores for individual documents \cite{sun2024chatgptgoodsearchinvestigating, pradeep2023rankvicunazeroshotlistwisedocument, pradeep2023rankzephyreffectiverobustzeroshot, ma2023zeroshotlistwisedocumentreranking, sun2025investigationpromptvariationszeroshot}. \cite{sun2024chatgptgoodsearchinvestigating} evaluates the performance of LLMs in RAG using prompting strategies and introduces RankGPT. To tackle the issue of context length limitations, it proposes a sliding window-based approach, breaking the primary prompt into smaller prompts and then merging the results of these smaller prompts. It then evaluates the model on various datasets to test the efficacy of its proposed solutions. To counter concerns that LLMs might have been trained on existing benchmark data, the paper introduces a new test set, NovelEval, with information post the knowledge cutoff date of GPT-4\cite{openai2024gpt4technicalreport}. Other models have also researched sliding window techniques, for example, \cite{rathee2025guidingretrievalusingllmbased} shares a SlideGar (Sliding-window-based Graph Adaptive Retrieval) algorithm to use adaptive retrieval.

\cite{bai2024finetuninglargelanguagemodel} investigates the challenges in the recommendation system due to the disparity between the pre-training LLMs and the specific requirements of recommendation tasks. It introduces a Direct Multi-preference optimization framework and fine-tunes the model by incorporating multiple negative samples and optimizing the model to increase the probability of correct predictions and decrease incorrect predictions. \cite{ma2023finetuningllamamultistagetext} uses supervised fine tuning  (SFT) on LLaMA\cite{touvron2023llamaopenefficientfoundation} for recommendation systems and introduces two models, RepLLaMa and RankLLaMa. RepLLaMa encodes queries and documents into a vector representation and computes the relevance using the dot product, while RankLLaMa performs pairwise ranking of the inputs. In contrast to SFT, \cite{zhang2024twostageadaptationlargelanguage} conducts training on a large weakly supervised corpus and proposes a two-stage training framework to adapt decoder-only LLMs to text ranking progressively using  Continuous Pre-Training(CPT)  followed by SFT. The base LLM undergoes CPT on a large-scale, diverse, weakly-supervised text-pair dataset from various sources. This stage uses the Next Token Prediction (NTP) objective to initially orient the LLM towards different aspects of relevance inherent in the text pairs. The model is fine-tuned using high-quality supervised ranking data. It proposes a new optimization strategy using a ranking loss based on the generation probability of the entire query given the document (P(q|d)), calculated via NTP, rather than relying on a last-token representation.

LLMs have also been proven to be effective text rankers when used in pairwise ranking settings. \cite{huang2024instuprinstructionbasedunsupervised} uses a pairwise approach where LLM is given a query and two passages to determine the relevant passage. All the comparisons made create a final score for the passage. \cite{luo-etal-2024-prp} also shares a pairwise prompting approach called PRP-Graph, which leverages LLMs’ output probabilities for target labels, transforming these outputs into numerical scores for enhanced text ranking.

Researchers have also used model distillations to create open-source models. \cite{pradeep2023rankvicunazeroshotlistwisedocument}, a LLaMa-based model is the first fully open-source LLM capable of performing high-quality listwise reranking in a zero-shot setting. It is trained in zero-shot settings using RankGPT\_{3.5} as a teacher model using 100K queries from the MS Marco dataset. However, \cite{pradeep2023rankvicunazeroshotlistwisedocument} still lags behind the state-of-the-art RankGPT4 in effectiveness. Thus, \cite{pradeep2023rankzephyreffectiverobustzeroshot} introduces RankZephyr, another open source model which bridges the gap and, in a few cases, outperforms RankGPT\_{4} with orders of magnitude fewer parameters. \cite{zhang2023rankwithoutgptbuildinggptindependentlistwise} studies how to construct effective GPT-free listwise rerankers based on open-source LLM models, where they used Code-LLaMA-Instruct\cite{rozière2024codellamaopenfoundation} for their experimental analysis.

In addition to these techniques, research has considered prompt engineering for reranking. \cite{jin2024apeerautomaticpromptengineering} shares a novel approach to reduce human effort and unlock the potential of prompt optimization in the reranking approach by generating refined prompts through feedback and preference optimization. \cite{wu2024passagespecificprompttuningpassage} integrates a soft prompt with a passage-specific embedding layer to form a new learnable prompt module, then concatenates with the embeddings of the raw passage to serve as new input for the LLMs. 

\section{Conclusion}

This survey tracked the evolution of reranking algorithms from foundational Learning to Rank (LTR) methods, which established principles for optimizing result order, through the transformative impact of deep learning that enabled sophisticated pattern capture and list-wise modeling. We then examined how knowledge distillation became crucial for deploying complex models efficiently, and finally, explored the emerging role of LLMs, which offer potential for deeper semantic understanding. However, current rerankers, particularly those leveraging deep learning and LLMs, grapple with challenges such as substantial computational overhead, making real-time deployment in high-throughput systems difficult. Furthermore, issues related to bias amplification and the need for extensive, high-quality labeled data remain significant hurdles in achieving robust and fair reranking performance across diverse queries and contexts. This development demonstrates an ongoing push for more efficient and contextually sensitive result ordering, with future research probably concentrating on combining these approaches and tackling the real-world difficulties of implementing sophisticated rerankers.

% \bibliography{surveyReranker}
\printbibliography




% LNCS Homepage, \url{http://www.springer.com/lncs}. Last accessed 4
% \end{thebibliography}
\end{document}